\documentclass[10pt,floatfix]{iopart}
\usepackage{graphicx}  
\usepackage{times}  

\newcommand{\onlinecite}[1]{\cite{#1}}

\begin{document}
\renewcommand{\textheight}{7.48truein}

\title
[Structure Prediction for $d$(Al$_{70}$Co$_{20}$Ni$_{10}$)]
{Energy-based Structure Prediction for $d$(Al$_{70}$Co$_{20}$Ni$_{10}$)}

\author{Nan Gu\dag\ddag, M. Mihalkovi\v{c}\dag\S, and C. L. Henley\dag}

\address{\dag\ Dept. of Physics, Cornell University,
Ithaca NY 14853-2501, USA}

\address{\ddag\ Present address: Dept. of Physics, 
M.I.T., Cambridge MA 02139 [CORRECT THIS]}

\address{\S Permanent address:
Institute of Physics, Slovak Academy of Sciences, 84228 Bratislava, Slovakia.}

\def\Ao{{\AA{}}}
\def\subphase{{phase }}
\def\subphases{{phases }}

\def\bfreply#1 {{#1}}  
\def\CLH#1 {{\bf CLH: #1}}
\def\NAN#1 {{\bf NAN: #1}}
\def\MM#1 {{\bf MM: #1}}
\def\LIR {{\bf LIR}}
\def\OMITa#1 {{\sl #1}}
\def\OMITb#1 {{}}
\newcommand{\glue}{Star cluster}
\newcommand{\glues}{Star clusters}
\newcommand{\Dec}{13\AA{}D}
\newcommand{\Decs}{13\AA{}Ds}
\newcommand{\Decagon}{13\AA{}~Decagon}
\newcommand{\Decagons}{13\AA{}~Decagons}

\begin{abstract}
We use energy minimization principles to 
predict the structure of a decagonal quasicrystal - 
$d$(AlCoNi) - in the Cobalt-rich phase. Monte 
Carlo methods are then used to explore configurations 
while relaxation and molecular dynamics are used to obtain 
a more realistic structure once a low energy configuration 
has been found. We find five-fold symmetric decagons 
~12.8\AA{} in diameter as the characteristic formation
of this composition, along with smaller 
pseudo-five-fold symmetric clusters filling the spaces between the
decagons.
We use our method to 
make comparisons with a recent experimental approximant 
structure model from Sugiyama et al (2002).
\OMITb{The program we 
use takes interatomic pair potential calculated from 
Generalized Pseudopotential Theory and empirically 
determined lattice constants as inputs.  We also assume 
that the quasicrystal can be described by a Penrose tiling 
with properly placed atoms on the Penrose tiles.}

\end{abstract}



\section{Introduction}

This paper reports structural predictions for the
decagonal quasicrystal $d$(AlCoNi) in the Cobalt-rich 
(`basic Co') \subphase, of approximate composition Al$_{70}$Co$_{20}$Ni$_{10}$ 
and atomic density 0.068 atoms/\AA{}$^3$, using simulations
to minimize energy, using  the same methods (and codes) as previous 
papers by Mihalkovic, Widom, and Henley~\cite{alnico01,alnico02,alnico04}
on the `Ni-rich' \subphase; this constitutes a first test of 
the transferability of that approach to other compositions that
are described by other tiling geometries. 

The Aluminum-Nickel-Cobalt  alloy $d$(AlCoNi) is the best
studied and perhaps the highest quality of equilibrium 
decagonal quasicrystals. Decagonal $d$(AlCoNi)
has about eight modifications (which we call `\subphases' here), 
each existing in a tiny domain 
of the phase diagram.\cite{Ri96b,Tsai96,Ed96,Gr96,Ri98}.
The compositions `basic Ni' (around $\rm Al_{70}Co_{10}Ni_{20}$),
and `basic Co' 
are the best-studied \subphases, having
the simplest diffraction patterns.
(The other \subphases show superstructure peaks, indicating
modulations.)
Several crystal approximants related to `basic Co' are known~\cite{Gr98},
one of which has a solved structure~\cite{Su02}.

We find that (in contrast to the `basic Ni' \subphase), the framework of
`basic Co' at $T=0$ is a network of edge-sharing $12.8$\AA{} diameter 
decagons (placed like the large atoms in a ``binary tiling'' quasicrystal).
A strong (but not always simple) Co/Ni ordering is found. 
Our model reproduces most, but not all, of the atomic positions
in W(AlCoNi). (A complementary brief account of our results is in
Ref.~\cite{Gu-pucker05}, and a more complete account is in
preparation~\cite{Gu-PRB05}.)

\section{Methods}

Our main assumption is that the $d$(AlCoNi) structure is
well approximated by a stacking of equally spaced 
two-dimensional tilings built from the Penrose rhombi under
periodic boundary conditions.
The chief experimentally determined inputs 
were the values $c = $4.08\AA{} for the stacking period 
and $a_0 = $2.45\AA{} for quasilattice constant 
(edge length of small rhombi).
\bfreply{It should be noted that the actual period for the Co-rich \subphase is
generally believed to be $2c$; we used $c$ because (i) it is
the simplest starting point, and was used in the prior
study of the Ni-rich \subphase~\cite{alnico01,alnico02} 
(ii) so long as the atoms are fixed  on discrete sites
(see Section~\ref{sec:results}), it turns out that period 
$c$ is optimal even when a cell with more layers is allowed; 
(iii) even in our final, relaxed period $2c$ structure 
(Section ~\ref{sec:RMR}), most atoms do repeat with period $c$.}

\begin{figure}
\includegraphics[width=3.1in,angle=0]{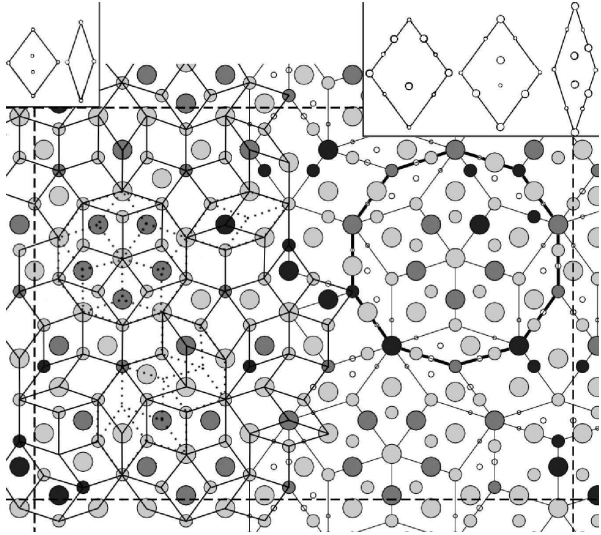}
\caption{The 32$\times$23$\times$4 unit cell. The left
half is subdivided into $a_0$ scale rhombi (in two ways in
part of the cell [dotted lines]). The right half of the cell is 
subdivided into $\tau a_0$ scale rhombi. \bfreply{The atomic
configuration shown was obtained from the $\tau a_0$ scale simulation.
(The atomic locations are also valid for the $a_0$ scale 
decoration, but configurations from $a_0$ scale simulations
almost always have more defective occupations.)}
Insets at top left and top 
right display the candidate sites on the $a_0$ and $\tau a_0$ scale 
tiles, respectively. Filled large circles depict atoms in one 
layer and filled small circles are those in the other layer. 
Black denotes Nickel, dark gray Cobalt, and light gray 
Aluminium. 
Unoccupied sites of the $\tau a_0$ scale tiling
are shown by very small empty circles;
those of the $a_0$ scale tiling are not shown. 
\bfreply{Four \Decagon~ motifs (described in Sec.~\ref{sec:decagon})
are present: e.g. one is seen decomposed into ten $\tau a_0$ edge rhombi
near the cell's upper right corner.}}
\label{fig:tiling}
\end{figure}

This framework of decorated tilings takes advantage of the fact
that all the known atomic structures of decagonals are
Penrose-tiling-like. Further, even in random tilings
Penrose rhombi admit ``inflation'' constructions that can relate
a tiling to another one using rhombi with edges enlarged by
$\tau = {{(1+\sqrt 5)}/{2}} \approx 1.618$ 
(the golden ratio), which gives a convenient way to 
build a chain of connections from the atomic level
to large tiles so as to describe $d$(AlNiCo) on large scales.
Fig.~\ref{fig:tiling} includes two ways of 
subdividing a unit cell into a tiling of rhombi (with edges
$a_0$ and $\tau a_0$, respectively).


We also assume atomic pair potentials between 
Al, Co, and Ni as derived ab-initio using Generalised Pseudopotential 
Theory~\cite{Mo97} (GPT),
but modified using results from 
{\it ab initio} calculations to 
approximate the sums of the 
omitted many-body interactions when transition metals (TM) 
are nearest neighbours.
[Fig.~1 of Ref.~\onlinecite{alnico01}
plots these potentials.]
The potentials are cut off by a smooth truncation past 7.0\AA{}. 

We know~\cite{alnico01,alnico02} that the strongest interaction
is the Al-TM first potential well. The number of Al-TM neighbours
is maximised and, as a corollary, that of TM-TM nearest neighbours is minimised
(but TM-TM second neighbours, at $\sim 4$\AA{} separation, are common).
However, this scarcely constrains the structure since the weak 
Al-Al potential allows enormous freedom in placing Al.
In practice, one finds in Al-TM quasicrystals that a rather
rigid TM-TM network forms,  with separations near to the
strong second minimum of the TM-TM pair potential; the Al atoms 
fill the TM interstices in a more variable fashion.
Subtle details of the density and composition, as well as the 
small differences between Co and Ni in the pair potentials, 
decide which structure optimises the energy.  
We understand only a few of these factors in detail for the
`basic Co' phase, so in this paper we will only describe the 
atomic structures we found and do not attempt a microscopic 
rationalization of them.

Our program uses Metropolis Monte Carlo (MMC) 
simulation to perform atom-atom and atom-vacancy 
swaps within a site list placed on the Penrose tiles. 
The first stage simulations include an additional degree of 
freedom in the form of `tile flips.' These 
`tile flips' obey the MMC algorithm probabilities 
and conserve tile number, atomic species number, 
and the collective outline of tiles rearranged. 

Our procedure is to anneal
from a high temperature $\beta \approx 4$-10 eV$^{-1}$
to a low temperature at which most degrees of 
freedom are frozen out, $\beta \approx 20$ eV$^{-1}$,
using a step size $\delta \beta \approx 0.5$-1 eV$^{-1}$.
(Here $\beta \equiv(k_{B}T)^{-1}$.) At each temperature step, 
approximately 2000 atomic swaps per site are attempted. 

\OMITb{
After annealing, we pick out the lowest energy configuration for 
analysis. It is important to point out that we do not average 
over emsembles, nor do we select the end result of an annealing 
process, but instead single out the lowest energy configuration. 
This procedure may give surprisingly stable configurations even at 
high temperatures.}

\OMITb{
Periodic boundary constraints permit only a 
discrete sequence of simulation cells. Particular cell sizes are 
especially favourable since they permit a tiling that is close to 
having five-fold symmetry (in the frequency of the various 
orientations of rhombi or other objects in the tiling.) }

\OMITb{There are more subtle details in the use 
of a tiling that will be elaborated on in a future paper.}

\OMITb{The steps described above apply to configurations that occur on a 
discrete site list, which will be the focus of our paper.}

\OMITb{ The 
relaxation portion of the program receive atomic configurations 
and performs the relaxations to 0 Kelvin. The molecular dynamics 
portion sets the atomic configuration to a Maxwell velocity 
distribution and anneals down to a low temperature. }

\subsection{Multiscale Procedure}

The strategy of our series of simulations loosely copies that 
reported previously~\cite{alnico01,alnico02} on the Ni-rich 
\subphase of $d$(AlNiCo), and can be thought of as a sort of 
multiscale modelling. It proceeds in the following stages:

\begin{itemize}
\item [1.]
Begin with Monte Carlo simulations allowing atom swaps and 
`tile-flips' with a tiling of relatively small Penrose rhombi;
note prominent features and formations that recur in all the 
low energy structures from a sufficient number of independent runs.

\OMITb{The unit cell is chosen to be as large a possible limited by 
a reasonable computation time. A large unit cell is necessary 
for us to be fairly certain that we can witness the formations 
unaffected by periodic boundaries.}

\item [2.]
Take the observations from the small-scale tiling and promote 
them to rules: i.e. make the formations inferred in the first 
stage into fundamental objects of a larger scale tiling. 
Unused (or underused) degrees of freedom are removed. 

\item [3.]
Perform new Monte Carlo simulations on the larger-scale 
tiling to search for new low energy structures. This larger-
scale tiling is generally more efficient at finding lower
energies because of the reduced degrees of freedom.

\item [4.] 
Verify that the degrees of freedom removed in stage (2) were 
indeed unnecessary by restoring some of them, judiciously.

\item [5.]
Repeat for larger and larger scale tilings. An eventual goal 
is to extract a `tile Hamiltonian'~\cite{Mi96a,Co98a} which 
is the bridge to modelling long-range order and diffuse scattering. 
That means ascribing the atom-atom energies to an effective
interaction among neighbouring tiles, so that tiles are the only
remaining degree of freedom. 

\end{itemize}

It is also necessary to perform relaxations/molecular dynamics 
to escape the biases which may be introduced by the discrete 
site list used up through stage (4).

\section{Results}
\label{sec:results}

\subsection{Small-Scale tiling (edge $a_0=2.45$\AA{})}
\label{sec:decagon}

The first stage of our tiling simulations uses small Penrose rhombi
that have $a_0=2.45$\AA{} edges.
\OMITb{The rhombi are decorated as shown in Fig.~\ref{fig:tiling}
by candidate sites (meaning they are free to be occupied or not
by an atom.)}
The unit cell simulations had dimensions $32.01$\AA{}
$\times 23.25$\AA{}$ \times 4.08$\AA{}, which we shall call the 
32$\times$23$\times$4 cell, shown in Fig.~\ref{fig:tiling}. 
The structure is taken to repeat 
after two layers of the Penrose tiling.
At 0.068 atoms/\AA{}$^3$, there are about 200 atoms in 
the unit cell. 

\OMITb{The atomic sites are placed at the
vertices of each rhombus and also on the interior of the 'fat-
rhombus' (the one with 72$^{\circ}$ and 108$^{\circ}$ angles)}

Our low energy configuration generated at this level showed
distinct rings of 5 TM and 5 Al atoms surrounding an Al atom 
and surrounded by an 10 Al atom decagon with an edge length 
of $a_0$. These 21-atom motifs were evidently energy 
favourable. Linear regressions of the count of these 
21-atom clusters  versus
energy revealed a noisy, but consistent correlation of 
$\approx -1.0$ eV per 21-atom cluster.

\OMITb{Just a note on the following para: 
Composition of the 21-atom motif is 5Co+16Al, ei 76\% Al.
So it is quite likely you cannot have more of these clusters
just by the composition constraint... though it is true
the outer shell is pure Al, and if this was shared by more
clusters, the TM content would be higher than those 24\%.
}

\OMITb{
``After the initial runs, it appeared that the number of 21-atom clusters
was not to be maximized. In the low-energy configurations, the 
Al$_{10}$ decagons typically were not adjacent to each other, and 
a typical maximum of four 21-atom clusters formed when it is feasible for
at least six to appear without conflict.
We ran additional MMC simulations with an initial configuration
taken from the low-energy runs. Low-energy results of these 
simulations showed that larger decagons emerged
with an edge length of $\tau a_0 \approx 4.0$\AA{} 
(hence a diameter of $\sim 12.8$\AA{}). ''}

However, the ideal structure did not simply maximsze
the number of 21-atom clusters: though they may be
placed adjacent to pack six clusters per cell, that was
typically not observed in the low-energy configurations. 
Instead, further MMC annealing (starting from a 
low-energy configuration) found that a larger cluster 
emerged as the characteristic motif at this composition:
a decagon with an edge length of $\tau a_0 \approx 4.0$\AA{} 
(hence a diameter of $\sim 12.8$\AA{}), 
which we will call `\Decagon' (\Dec).
This object contains the
21-atom motif at its centre, ten TM atoms at its vertices,
and Al atoms irregularly interspersed along and within its edges.

We use the following 
nomenclature for the components of \Dec, from its
centre outwards. 
The centre of the \Dec~ is an Al atom.
The 5 TM 5 Al ring is known as the `first ring.'
The TM atoms in the first ring are in the same layer as the central
Al atom.
The 10 Al ring is known as the `second ring.'
The 10 TM decagon and the Al atoms along and within the edges are 
collectively known as the `third ring.'
Miscellaneous Al sites occur between the second and third rings, 
which  we designate collectively as the `2.5 ring.' 

In the \Dec, the first ring TM sites are mainly occupied by Co 
and the third ring (decagon vertex) TM sites also tend to Co occupation, 
but there is a very strong dependence on the local environment that distinctly
favour Ni in certain sites.
[This became clear at the second stage of simulation, 
Subsec.~\ref{sec:inflated}.]
The Al positions of  third-ring and 2.5 ring Al atoms are closely 
correlated and the rules were not resolved from the 2.45\AA{} tiling
simulation.


A second type of motif fills the 
interstices between the \Decs: it consists of 
a 10-atom ring and an Al at the centre, 
much like ring 1 in the \Dec~ cluster, except this kind
of motif has mixed Al-TM occupations for the 5 candidate TM sites. 
In general, about 3 sites out of the 5 candidates are filled
with TM atoms, preferably Ni.
We name these 11-atom clusters ``\glues'';
a few of these are contained in 
Fig.~\ref{fig:tiling}.

\bfreply{Perfect examples of either cluster were rarely observed in the 
first (2.45\AA{}) level of simulation: their decorations were 
induced by a `consensus' or average over many defective examples.
Thus, the rules for Al/TM occupation on the \glue~ TM sites, 
and for Ni/Co occupation on all TM sites, were still incompletely known 
from  this stage.}

\OMITb{These answers to there problems seemed difficult 
to obtain at best with more MMC refinements on the $a_0$ 
tiling. but we've obtained a sufficient amount of 
information to eliminate many degrees of freedom.}


\subsection{Inflated Tiling (${\tau}^{3} a_0 \rightarrow \tau a_0$)}
\label{sec:inflated}

\OMITb{The $a_0$ tiling results were sufficient information
to allow us to formulate another level of description 
in which many degrees of freedom were eliminated.}

In keeping with our `multiscale' methodology, we reassessed
the degrees of freedom necessary for the next level of simulation.
The dominant geometric object is a decagon with edge $\tau a_0\approx 4$\AA{}, 
but we found it economical to represent its decoration by candidate
sites using a thin 
4\AA{} rhombus and the second fat rhombus shown in the right inset
of Fig.~\ref{fig:tiling}; similarly, the \glue~ decoration is represented
using the first fat rhombus.  

We assumed that the minimum-energy 
structure has a maximum density of \Dec~ clusters.
A Monte Carlo simulation was carried out
(admitting only tile flips) 
on the ensemble of $\tau a_0$ rhombus tilings
using an artificial tile Hamiltonian
favouring the creation of nonoverlapping `star decagons' (of edge $\tau a_0$)
containing a fivefold star of fat rhombi.~\footnote{
Star decagaons were similarly maximized in Ref.~\onlinecite{Hen98}, 
except they could overlap in that case.}
It was found that the ground state configuration was 
always a `binary tiling'~\cite{La86,Wi87} of rhombi
with edge length ${\tau}^{3} a_0 = $10.4\AA{} with 
the `large atom' and  `small atom' vertices replaced
by \Decs~ and \glues, respectively.
\bfreply{(The constraints on possible decagon separations imposed by
building them from $\tau a_0$ tiles  are consistent, in fact, 
with those due to the potentials~\cite{Gu-PRB05}.)}

\OMITb{broken into 5 thin and 5 fat $\tau a_0$ edge length Penrose rhombi. }

We {\bf designate} a fixed site list on the $\tau a_0$ rhombi, 
as shown in the right half of Fig.~\ref{fig:tiling}, designed
to match the most frequently occupied sites observed in the
low-energy states from out $a_0$ scale simulations. 
This tiling, unlike the $a_0$ tiling, consists of only {\it one} layer 
of Penrose tiles, each of which has site decorations in 
{\it two} layers spaced 2.04\AA{} in the $c$ direction. 
Note how the two versions of the fat Penrose rhombus have
different site lists. 
We use the 
32$\times$23$\times$4 cell, as before so we may compare the energies
for systems with exactly the same density and composition.

\OMITb{ DONE REMOVED The \Dec~ has a natural {\sl five-fold} $\tau a_0$ edge rhombus 
decomposition, which is shown in Fig 1. 
This decomposition requires 
that the local tiling must be in the form of a decagon. Thus, we 
take as an {\it a posteriori} rule that the number of \Decs~ 
are maximized on a given tiling. Thus, an initial tiling is manually 
'flipped' until an optimum tiling configuration is obtained. The 
optimum tiling in the 32$\times$23$\times$4 unit cell contains four 
\Decs~ configured in a herringbone lattice. }

At this second stage,
our MMC runs had purely lattice gas moves on
a fixed tiling (no tile flips).
These were able to find lower energy 
configurations at a much faster rate than the first-stage ($a_0$ scale)
simulations at the same composition and atomic density, 
due to the much decreased site list.
Hence, the occupations of many types of sites were discerned
with greater accuracy, such as the TM sites in the \Decagon~
(already described in Subsec.~\ref{sec:decagon}). Furthermore the 
tendencies of more variable sites (Al in rings 2.5 and 3, and 
TM in the \glue), which will be described in a longer paper~\cite{Gu-PRB05},
started to come into focus.
Finally, despite having {\it fewer} allowed configurations, 
the $\tau a_0$ tiling simulation consistently found {\it lower} 
energies, which serves as our post hoc justification for assuming
\Dec~ clusters in the binary tiling geometry and the reduced site list.
\footnote{
In additional tests, certain candidate sites that are absent in the 
inflated tiling were restored systematically to the $\tau a_0$ tiles, 
and it was verified this had a negligible effect.}

\OMITb{We implement the tiling in this fashion because we note that
certain sites of a Penrose rhombus are always or rarely used 
depending on the location of the rhombus. The rhombi can be 
distinguished and pieced together in a consistent manner by adding 
an additional degree of freedom in the labeling of each rhombus 
vertex.  The details of this additional degree of freedom are more 
subtle and pending a longer future paper. }

\OMITb{The tile flips are disallowed because
we have optimized the tiling for what we believe to be the lowest 
energy structures. Although we have implemented what may be an 
unfair amount of constraints, we still allow an important degree of 
freedom, namely the chemical occupation (incluing vacant) of each 
site. }

\OMITb{
\NAN{Maybe I should leave this paragrah out...thoughts? Update
I'm going to take it out. The next one too.}
we test the $\tau a_0$ scale tiling on nearby compositions,
densities, and unit cells. We find that although the details depend
very much on the specifics of the parameters used, that \Dec~ gross
structure is robust under these parameter changes. Thus, we can also 
infer that the \Dec~ maximization rule holds weight at these nearby
points in parameter space. The existence of the \glue~  is also
guaranteed, but is pretty much fixed by our choice of tiling. We may 
expect a portion of the phase diagram to be dominated by \Decs, but 
must exercise caution in any implementation.}

\OMITb{
Further MMC testing to pin down the exact occupation of the sites
deemed uncertain in the $a_0$ scale tiling are partially successful
at best. We have found that the TM occupation within any \Dec~
tends to be entirely Co, while the Ni atoms are forced off to the
\glues. There are a set of rules that the configurations
tend to follow, but these rules are dependent on the tiling, unit cell
size and composition. These rules are subject to change upon exploration 
of new unit cells, compositions, and atomic densities and a comprehensive
list will be pending a longer future paper.}

In the prior case of the `basic Ni'  \subphase, the 
decoration model~\cite{alnico01} 
was implemented as a deterministic tiling.
A marked contrast in the present `basic Co' case is that
we cannot impose a simple rule that fixes the chemical occupation 
of each site. On certain TM sites, it is difficult to resolve the
occupation (Co/Ni); certain Al sites are also variably occupied 
(Al/vacant) depending on the environment. 
Presumably, with a sufficiently thorough understanding of our 
model, we could formulate a deterministic decoration rule on 
the binary tiling. It is possible, however, that the energy differences 
among some competing structures are too small to be visible
in a reasonable simulation, or to influence the real properties 
at any accessible temperature.

\OMITb{Can't this be moved to the Discussion?
It is possible, however, that the energy differences favouring this 
resolution will be so small that we would  be unsure whether, e.g., the 
rigid-site-list and MD-relaxed energies would agree in predicting 
this structure. This is one of the main pitfalls of the lattice-gas
model that we have used.}

\OMITb{In particular, though the TM atoms on the 
boundary of \Decs~ should be idealized as Co, there are special 
environments in which they are converted to Ni.  We also find that 
the 2.5 and third rings as well as the \glues~ have occupations 
that are very much dependent on their surroundings. These degrees of 
freedom interact with the tiling geometry, as well as each other.}

\subsection{Cluster orientation}
\label{sec:orientation}

The physical accessible candidate sites in the $\tau a_0$ decagons
(with our decoration) in fact have a $\overline{10}$ point group symmetry, 
which is not broken by the binary tiling geometry.  The fivefold
symmetric \Dec~ cluster breaks this symmetry, by the layer on which 
the central Al sits and the Al/TM alternation on the first ring. 
Thus, a major obstacle to writing a deterministic decoration is 
that the relative orientations of \Decs~ must be specified, 
which depends on subtle interaction energies between them.
Experimentally, $d$(AlCoNi) with our Co-rich composition was
observed to order in a fivefold symmetry, wherein all the Decagons
are oriented the same way~\cite{Ri96c,Li96}.

As a test, we compared the lowest-energy configurations 
resulting from MMC simulations on identical $\tau a_0$ tilings
in which neighbouring \Decs~ were forced either to have always 
identical or always opposite orientations. [This is controlled
by the orientation of the decagon of $\tau a_0$ rhombi in
the tiling that gets decorated.]
we evaluated the minimum-energy configurations from
$\sim 60$ independent runs of each type.
We used the 32$\times$23$\times$4 unit cell, using a fixed composition 
Al$_{70}$Co$_{20}$Ni$_{10}$, but repeating the tests for a 
a series of atom number densities $n$.
\OMITb{MM:For experimentalists, it would be interesting to know
the consequence of this on the space group.}

The energy difference between these two orientations
was density dependent. At $n\approx 0.068$ atoms/ \AA{}$^3$ the 
two orientational patterns are practically degenerate
(though with visibly distinct atom configurations.)
From $n=$0.068 \AA{}$^{-3}$ to 0.074 \AA{}$^{-3}$ --
a range which includes the most realistic compositions --
the pattern with uniform orientations has the lower energy,
by up to $4\times 10^{-3}$eV/atom, but
this difference disappears again at $n \approx$0.074 \AA{}$^{-3}$.

We find this energy difference arises from different ways of 
filling the possible TM sites in the respective orientation schemes. 
In the uniform-orientation scheme, 
as the TM density is raised (as part of the total density),
TM atoms fill the ring 2.5 in the \Decagon~
before they exceed $\sim {{3}\over{5}}$ filling in the \glue~ TM sites;
whereas in the alternating-orientation case, the
ring 2.5 is filled to a lesser degree with TM while the \glue~ TM sites
become overfilled. [In Ref.~\onlinecite{Gu-pucker05}, we gave an explanation
how these effects could favour {\it alternating} orientations.] 
But the results just presented are valid only for the 
(unphysical!) case of atoms confined to hop on fixed ideal sites.
When displacements are allowed, so that many atoms ``pucker'' away from
the flat layers (see Sec.~\ref{sec:RMR}), then the uniform arrangement
is preferred more robustly.

\OMITb{We find a slight energy difference favouring antiferromagnetic 
configurations exists when tested upon the fixed site list. 
Under RMR, the energies of ferro- and anti-ferromagnetic 
configurations merge. Based on the results of our RMR tests, 
we believe that any nearest neighbour interactions are mediated 
by puckering effects at least as influential as the effects 
of the discrete site list.}


\OMITb{The 4.08\AA{} periodic structures that we study do not exhibit any significant form of out of layer relaxation (puckering). 
Hey, it could, it just doesn't.  Symmetry guarantees that positions
in the layers are stationary with respect to moving an atom up or down.
But it could be a saddle point in the z direction; not to mention the
possibility that a group of atoms would collectively move up or down
together.  There is just something about the potentials that adds up
to favour positions right in the layer...}

\OMITb{The diffuse diffraction patterns 
\cite{Fr00a,Fr00b} of the 'Co-rich' phase suggest local doubling of the 
period to 8.16\AA{}. In the case of such a doubled cell, we may 
look for any stabilizing effects 
that may result from puckering.}

\OMITb{Fixed site list 8.16\AA{} periodic structures tend to have a remarkable
approximate 4.08\AA{} periodic symmetry. Sites in each 4.08\AA{} sub unit 
cell tend to have the same chemical occupation not just between Al and TM, 
but also between Co and Ni. This 4.08\AA{} approximate symmetry makes 
puckering an approximate spontaneously broken symmetry. When the discrete
site list is relaxed, the atoms in certain layers pucker one way or 
another, creating `flat' and `puckering' layers. 
Puckered Al atoms (allowed to anneal sufficiently via RMR) can also intersperse
themselves with three Al atoms per 8.16\AA{}.}

\OMITb{CLH:The initial condition is practically a stacking in terms'
of occupations, so it seems the puckering
does break a symmetry.
MM:
Note that there is another type of symmetry-break that we dont
tackle: doubling of the unit cell for W-phase along 40A in-plane
lattice vector (natural approximant size is 20A, ei half).
In the W-phase, the `flat' layers are actually 4.08A periodic,
and the puckered layers are related by the (0 1/2 1/2) centring
(in 40 x 23 x 4.1 A cell).
So I think from this simulation we just get the small, `local' part
of the symmetry-breaking and puckering. 
The stronger part (as seen in real W-phase) of
the symmetry-breaking requires correlations between larger blocks
of atoms. I think this is where our method fails, because we
would need simultaneous action of puckering displacements plus
a kind of reshuffling to discover this.}

\OMITb{Under relaxations or RMR applied to an 8.16\AA{} structure
 we once again find the TM atoms mostly 
immobile with negligible puckering. 
Most Al atoms that pucker 
are found in the 2.5-th and third rings. These two rings act in concert
to lower the energy and should be considered together upon relaxation. The
details are complicated and will be discussed pending a longer future paper.}

\section{Beyond the Discrete Site List}
\label{sec:RMR}

\OMITb{
We are capable of removing the tiling and site list constraints and 
performing relaxation and molecular dynamics (MD) on any given 
configuration we obtain (usually the lowest energy one).  These effects will only 
be touched upon in this paper.}

Up to this point, we reported simulations using fixed 
atom site positions. We can make our configurations more realistic 
by subjecting them to relaxation and molecular dynamics (MD). The 
removal of the tiling and discrete site list leads to effects such 
as out-of-layer ``puckering'' and so-called `period doubling,' whereby
certain atoms relax into positions that violate the $c$ period of
a single unit cell, but instead are periodic with respect to a 
unit cell with $c'=2c =$ 8.16\AA{}. 
Exactly such distortions are familiar in the structures of
decagonal phases and approximants.

Our standard cycle for such `off-site' studies is a relaxation to a local
energy minimum (i.e. to 0 K), followed by an MD cycle
beginning at $\sim$600 K and ending at $\sim$50K in increments of 50K. 
The MD results are then relaxed again to 0K. This protocol will be denoted 
Relaxation-MD-Relaxation (RMR).
Upon relaxation to $T = 0$, we find that the TM atoms are quite immobile
and move only slightly. The Al atoms, however are subject to displacements as
large as ~$1.5$\Ao{}. After molecular dynamics and re-relaxation to $T = 0$,
we find a few Al atoms be further displaced, but to similar sites
so that no systematic difference is apparent in the overall pattern.

\subsection{Comparison with Experimental W-AlCoNi}

Major diffraction-based structure models were available for
`basic Ni' $d$(AlNiCo), to which 
the simulation predictions could be compared, but no such 
model exists for the `basic Co' $d$(AlCoNi).
However, the structure of the `W phase' crystal approximant of `basic Co' 
has been determined by Sugiyama {\it et al}~\cite{Su02}, and
we may apply what we have learned so far  -- in particular
the period doubling and puckering -- to predict its structure.
[This study has inspired a more detailed modelling of W(AlCoNi),
based on ab-initio energies rather than pair potentials~\cite{Mi05}.]

\OMITb{using the SIR97 package, which employs 
direct methods and least-squares-Fourier methods to extract atomic positions.}
Using exactly the same tiles and interlayer spacing as in our other
simulations, 
our unit cell is $23.25$\Ao{}$ \times 39.5606$\Ao{}$ \times 8.158$\Ao{}, 
which differs by less than 1\% 
from the experimentally determined lattice constants. 
The approximate atom content implied 
by Sugiyama's structure solution and Co:Ni ratio 
is Al$_{385}$Co$_{113}$Ni$_{38}$, 
with uncertainties arising from the partial and mixed occupations.
Adopting this composition as our ideal, we have
Al$_{71.8}$Co$_{21.1}$Ni$_{7.1}$ at a density
0.0714 \AA{}$^{-3}$.

We applied our mock Hamiltonian from Sec.~\ref{sec:inflated}
to generate a tiling consistent with the ${\tau}^3 a_0$ 
binary rhombus tiling. The optimum configuration is unique 
(modulo symmetries) and has four \Decs, in an arrangement
which turns out to be the same as observed in 
in W(AlCoNi). Cluster orientation comparisons
were like performed like those of Subsec.~\ref{sec:orientation}, 
but using the relaxed energies (without MD).
\OMITb{CLH asks: Relaxed, not RMR? [orig. text said just "relaxed".
Need to check!]}
We found that, between the configurations with alternating and uniform cluster 
orientations, the latter had a lower relaxed energy,
in accordance with the experimental W-AlCoNi structure. 
The atomic locations generated on this unit cell using the 
4\AA{} rhombus decoration site list
is approximately consistent with the experimentally determined sites. 

\begin{figure}
\includegraphics[width=3.1in,angle=0]{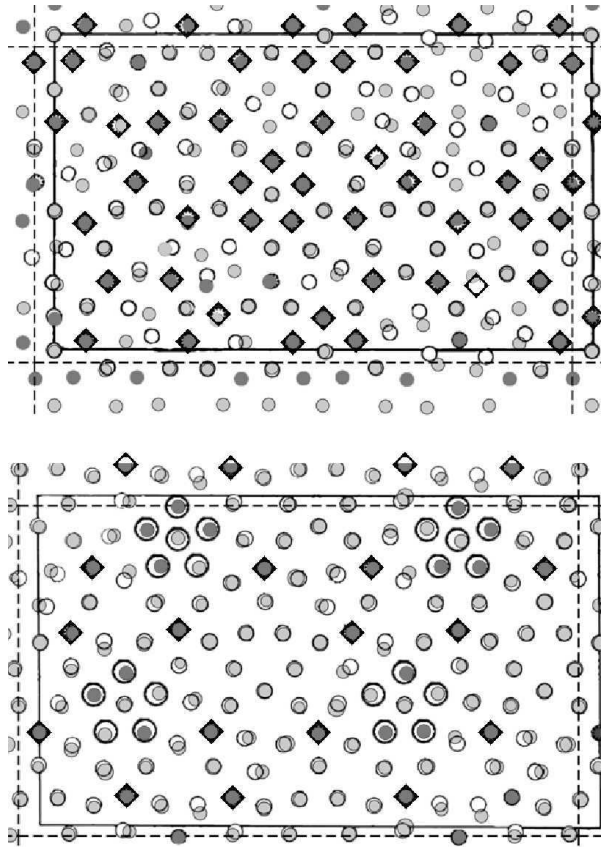}
\caption{a) The top figure is one of the mirror layers of W-AlCoNi. 
b) The bottom figure is the puckered layer of W-AlCoNi. 
The empty circles are experimentally determined atom sites, the 
filled circles are those extracted from our simulations. Dark gray 
filled circles are TM, light gray filled circles are Al. Empty circles 
with 'teeth' are TM, smooth empty circles are Al.
The large empty circles are experimental atomic sites with 
mixed Al-TM occupancies. The size of the simulated (filled) atoms 
may vary due to the visual scheme depicting 
distance in the $c$-axis dimension.}
\label{fig:WA}
\end{figure}

Our MMC simulations are able to capture the gross features of 
W-AlCoNi with excellent accuracy. Since this approximant is not far from 
the `basic Co' decagonal composition, 
we are not surprised to find \Decs~ and \glues~
as the major motifs, arranged in a manner consistent with our binary
and $\tau a_0$ tilings, consistent with the observed W-phase. The
differences in the exact atomic locations lie mainly within the highly 
context dependent 2.5th and 3rd rings. We find that the increase in density 
from our original simulations results in a more dense occupation of these 
rings. Atomic configurations found after RMR are in even better agreement. 
Fig.~\ref{fig:WA} compares one layer from experimental data to our RMR
results; the match in the other layers are equally good.

However, a significant number of Al atoms in our RMR picture 
are in disagreement with the experimental refinement.
because certain puckered atoms are not present in our simulations.
The top right of Fig.~\ref{fig:WA} 
contains a pentagon of displaced Al atoms centred upon 
the missing Al atom. These extremely puckered Al atoms are perhaps located
in energy minima too far away for our MD program to traverse. 
The existence of such atoms is one of the greatest flaws 
for beginning with a discrete site simulation.

\OMITb{
Another difference between our simulation and experiment lies in the exact
occupation of the \glues. Our simulations seem to show a few 
configurations of TM atoms that appear equally favourable while the experimental
\glues~ lie in fixed configurations. Once again the energies that
decide this effect may be dominated by puckering.}

\section{Conclusion}

We find that in the approximate composition (Al$_{70}$Ni$_{10}$Co$_{20}$,
the structure of $d$(AlNiCo) is dominated by the formation of
the \Dec~ clusters and the \glues~ which complement them. 
We find that the gross features are robust under variations
of the composition and densities of $\sim 4$\%; 
these will strongly affect a few details of the sites
and we have not established the correct answers for those atoms.
Our method of tile decorations, and successive reductions of the
degrees of freedom, affords an enormous speedup compared
to brute-force molecular dynamics, which would get stuck in
glassy configurations, at the price of a few plausible assumptions.

By applying our results to the approximant W-AlCoNi, we 
demonstrated both the validity and the pitfalls of our approach. 
The pitfall lies in that fact that our explorations are
conducted using a fixed site list in two atomic layers,
whereas the real structure has four layers and some atoms
are strongly puckered out of them. Even though most puckered
atoms may be represented as relaxations from ideal sites,
the optimal puckered structure might be derived from a
fixed-site structure of comparatively high energy that our
simulations would pass over.  Reliable answers can not be
obtained by pure numerical exploration, but seem to demand
some physical understanding of the puckering (and other
displacements), as we have begun to do in Ref.~\onlinecite{Gu-pucker05}.
The context dependences evident in  the `basic Co' show that
beyond a point, the errors in defining the pair potentials 
will surely exceed the energy differences due to some swaps
of species, which is a second pitfall that may be alleviated
only by checking against experimental phase diagrams.


\ack 
This work is supported by DOE grant DE-FG02-89ER45405;
computer facilities were provided by the Cornell Center for Materials
Research under NSF grant DMR-0079992.
MM was also supported by grant VEGA-2/5096/25 of the Slovak Academy
of Sciences.
We thank M. Widom for discussions.

\end{document}